\documentclass[11pt]{article}
\usepackage{amsmath,amssymb,graphicx}
\graphicspath{{figures/}}
\usepackage[numbers,super]{natbib} 
\usepackage[utf8]{inputenc}
\usepackage[T1]{fontenc}
\usepackage{geometry}
\usepackage{subcaption}
\usepackage{booktabs}
\usepackage{authblk}
\title{Experimental Demonstration of Plasmon-Enabled Monolithic Bragg Reflectors for Infrared Light via Inverse Design}
\author[1]{Mikołaj Badura}
\author[2]{Mikołaj Janczak}
\author[3]{Michał Rygała}
\author[3]{Tristan Smołka}
\author[1]{Adriana Łozińska}
\author[1]{Wojciech Dawidowski}
\author[4]{Paweł Piotr Michałowski}
\author[1]{Beata Ściana}
\author[3]{Marcin Motyka}
\author[2]{Tomasz Czyszanowski}
\affil[1]{Faculty of Electronics, Photonics and Microsystems, Wrocław University of Science and Technology, ul. Janiszewskiego 11/17, 50-372 Wrocław, Poland}
\affil[2]{Photonics Group, Institute of Physics, Lodz University of Technology, ul. Wolczanska 219, 93-005 Łódź, Poland}
\affil[3]{Laboratory for Optical Spectroscopy of Nanostructures, Department of Experimental Physics, Faculty of Fundamental Problems of Technology, Wrocław University of Science and Technology, Wybrzeże Wyspiańskiego 27, 50-370 Wrocław, Poland}
\affil[4]{Łukasiewicz Research Network --- Institute of Microelectronics and Photonics, 02-668 Warsaw, Aleja Lotnikow 32/46, Poland}
\date{January 8, 2026}
\usepackage{siunitx}
\newcommand{\um}[1]{\SI{#1}{\micro\meter}}

\newcommand{\bfum}[1]{\textbf{\num{#1}}~\si{\micro\meter}}
\newcommand{\percmcube}[1]{\SI{#1}{\centi\meter^{-3}}}

\DeclareSIUnit[quantity-product = ]\percent{\char`\%}
\newcommand{\pc}[1]{\SI{#1}{\percent}}
\renewcommand{\deg}[1]{\SI{#1}{\degree}}
\newcommand{\degC}[1]{\SI{#1}{\degreeCelsius}}

\newcommand{\omcm}[1]{\SI{#1}{\ohm\centi\meter}}

\newcolumntype{P}[1]{>{\centering\arraybackslash}p{#1}}
\newcolumntype{M}[1]{>{\centering\arraybackslash}m{#1}}

\usepackage{hyperref}
\begin{document}
\maketitle

\begin{abstract}
High-reflectivity mirrors in the mid-infrared (MIR) range are essential for next-generation optoelectronic devices but are still constrained by strain accumulation, poor thermal conductivity, and growth instability of thick multi-alloy stacks in conventional distributed Bragg reflectors (DBRs). We introduce plasmon-enabled DBRs (PE DBRs) based on modulation-doped monolithic InP, where plasmonic dispersion in highly doped layers provides a strong refractive-index contrast. Using inverse-design optimization targeting reduced free-carrier absorption and maximized reflectivity, we demonstrate that PE DBRs can achieve reflectivities approaching \pc{100}. Experimentally grown \um{14} thick InP PE DBRs exhibit up to \pc{99} reflectance with bandwidths reaching \pc{18} of the design wavelength. The monolithic, junction-free configuration ensures low resistivity and enhanced thermal performance, offering a scalable platform for efficient plasmonic mirrors in MIR photonics, with potential applications in photodetectors, light-emitting diodes and lasers.
\end{abstract}

\section{Introduction}
\label{sec:intorduction}
Mid-infrared optoelectronic devices have gained significant importance across various technology fields, including thermal imaging, free space communication, human-eye-safe LIDAR systems, and detection of toxic gases with strong absorption lines in the 4 to \um{10} wavelength range \cite{gordon2017}. In this range, interband cascade (IC) and quantum cascade (QC) active regions are widely employed in lasers (ICLs, QCLs) \cite{yang1995, faist1994, hlavatsch2022}, light emitting diodes (LED) \cite{schafer2019}, resonant cavity (RC) LEDs \cite{diaz-thomas2020}, photodetectors (PD) \cite{delga2020} and RC PDs \cite{green2004}. Surface emission in lasers and LEDs offers several advantages over edge emission, such as high optical power, wafer-level testing, two-dimensional integration enabling low beam divergence and near-Gaussian beam distribution in the far field \cite{iga2000}. PDs in a vertical configuration enable higher packing density, reduced footprint, increased responsivity, and low capacitance, resulting in faster response times \cite{chen2017}. For all surface emitting or detecting configurations, planar reflectors play a crucial role in their operation enhancing the quality factor of the lasers, RC LEDs, RC PDs, redirecting the light in LEDs and recycling the light in PDs. 

A conventional planar reflectors of high optical power reflectance utilize distributed Bragg reflectors (DBRs) that consist of numerous half-wavelength, optical thicknesses periods. Each period of a DBR in its most common configuration consists of two layers made of different materials with distinctly different refractive indices, negligible absorption, and thicknesses corresponding to one quarter of the design wavelength. The conduction band offset between materials of different refractive indices creates a Schottky barrier at their interfaces, which hinders current flow in electrically conductive DBRs. Moreover, the interfaces between layers in DBRs reduce the heat conductivity with respect to bulk material due to modification of phonon dispersion and phonon scattering \cite{gordiz2016}. These effects can be reduced by replacing the sharp interface between two different materials with a gradual compositional change. In such configurations, the unit block of the DBR has a half-wavelength optical thickness, and its constituent layers are no longer of quarter-wavelength thickness \cite{cheng2022}.

The epitaxial growth of semiconductor-based DBRs poses technical challenges due to mismatched lattice constants of materials with sufficiently high refractive index contrast. However prominent exception of $\mathrm{Al}_x\mathrm{Ga}_{1-x}\mathrm{As}$ exhibits acceptable lattice constants matching and high refractive index contrast, making it suitable for near and mid-infrared applications \cite{michalzik2013}. Furthermore, extensive research has been conducted on alternative materials that can achieve lattice matching with other systems, such as AlGaAsSb lattice matched to InP \cite{nakagawa2001} or AlGaSb lattice matched to GaSb \cite{sanchez2012} as well as AlInN lattice matched to GaN \cite{seneza2021} and AlInN layers matched to AlGaN both for visible spectral range \cite{feltin2006}. In the case of mid-infrared devices pose an additional challenge, even for lattice-matched materials, as the required total DBR thickness to achieve high reflectivity surpasses the critical thickness of the DBR stack. Exceeding critical thickness leads to defects and mechanical instability of the wafer. Conversely, DBRs composed of dielectric materials, which are not constrained by lattice-match requirements, face limitations such as high thermal resistivity and a lack of electrical conductivity. Recently, Li et al. \cite{li2024} provided an alternative approach of creating monolithic and homoepitaxial InP DBRs by electrochemical porosification of periodically distributed Si doped InP layers. Manipulation of the refractive index can be achieved through porosified InP containing voids of sufficiently small dimensions, which, for infrared light, form a homogeneous region with refractive index which value is averaged by volumes of InP and air. 

In the infrared spectral range, semiconductors exhibit pronounced variations in the refractive index, which originate from collective oscillations of free carriers (plasmons) induced by a time-varying electric field, as described by the classical Drude model. As will be discussed in Section~\ref{sec:reflection_one_layer}, these plasmons lead to a decrease in the absolute value of real part of the refractive index and a substantial increase in the absolute value of imaginary part. 

\begin{figure}[htbp]
\centering
\includegraphics[scale=1.0]{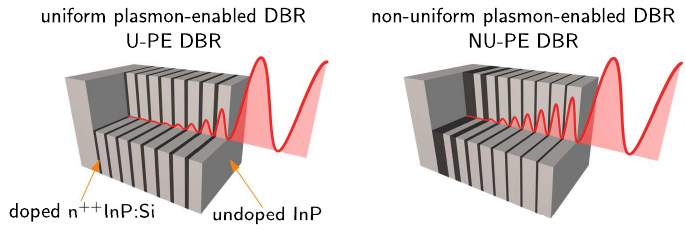}
\caption{Cross-sectional schematics of uniform (U-) and nonuniform (NU-) plasmon-enabled (PE) DBR composed of pairs of doped (dark grey) and undoped (light grey) layers of InP, implemented on an undoped InP substrate. The red curve represents the amplitude of light incident from air. Two configurations are considered in this work: U-PE DBR with equal thicknesses of the layers in each pair, and NU-PE DBR with the thickness of the layers modified in each pair as shown in the figure.}
\label{fig:DBR_visulization} 
\end{figure}

In this article, we utilize this variation in the complex refractive index induced by modulation of free electron concentration and propose a design for highly reflective mid-infrared DBRs using a monolithic InP structure with doping modulation, as illustrated in Fig.~\ref{fig:DBR_visulization}. The demonstration of the DBR with doping modulation serves as proof of concept for a broader approach, where any unitary or binary semiconductor material with modulation doping can function efficiently as a DBR for infrared light. Unlike multilayer DBR designs, this monolithic structure eliminates interface nonuniformities, stress-induced defects, and structural instabilities, thus removing the critical thickness limitation. Consequently, this configuration enables the growth of thicker layers required for IR applications and simplifies the epitaxial process by reducing the need for extensive preconditioning. Monolithic DBRs, composed of unitary or binary materials free from interfaces between layers with different lattice constants, also exhibit enhanced thermal conductivity compared to ternary or quaternary multilayer DBRs. The absence of such interfaces, which often induce abrupt bandgap discontinuities resembling Schottky junctions, further improves electrical conductivity, approaching values characteristic of bulk materials.

The DBR configurations that we discuss in this article are referred to as plasmon-enabled (PE) DBRs, as they utilize plasmon-induced changes in the refractive index to achieve high reflectivity in the mid-infrared range. The concept of plasmon-enabled (PE) DBRs was first proposed by Bergthold et al. in \cite{bergthold2022}. In their study, a PE DBR configuration composed of InAsSb/GaSb was designed for operation at a central wavelength of \um{4.2}. The design assumed that the thickness of the DBR layers adhered to the quarter-wavelength condition, $\lambda/(4n)$, where $\lambda$ is the central wavelength and $n$ is the real part of the refractive index of the DBR layers. This approach proved effective for wavelengths below the free carrier plasma resonance wavelength ($\lambda_\mathrm{p}$), approximately \um{6}. The optimized PE DBR achieved a maximum reflectivity of \pc{97}. 

In this article, we present an advanced approach to designing PE DBRs that goes beyond the simplistic assumption of quarter-wavelength thickness for the DBR layers. High doping layers, which create significant contrast in the complex refractive indices, enhance reflection at the interfaces between highly doped and undoped regions. However, absorption in the doped layers reduces the overall reflectivity of the PE DBR, making the design process more challenging and requiring the use of inverse-design approach. In this approach, the target optical response --- in this case, maximized reflectivity and minimized absorption --- is first defined, and the geometric parameters of the structure are then iteratively optimized to achieve this goal. Unlike conventional forward-design methods based on the quarter-wavelength rule, our method relies on numerical optimization of the doped and undoped layer thicknesses.

We propose two design schemes for PE DBRs. The first, referred to as the uniform PE DBR (U-PE DBR), uses identical layer thicknesses throughout each period. The second, the non-uniform PE DBR (NU-PE DBR) intentionally varies the thicknesses of the doped and undoped layers within each DBR segment to minimize absorption losses and maximize reflectivity (see Fig.~\ref{fig:DBR_visulization} for schematics). We numerically demonstrate that the reflectivity of NU-PE DBRs can approach nearly \pc{100} across a broad infrared range, including wavelengths beyond the free carrier plasma resonance. The findings of the numerical analysis are validated by the experimental realization of three InP NU-PE DBRs targeting $\sim \um{5}$, $\sim \um{7}$, and $\sim \um{9}$, with the latter achieving reflectivity as high as \pc{99.6}.
\section{Reflection from highly doped uniform layer}
\label{sec:reflection_one_layer}
\begin{figure}[htbp]
\centering
	\begin{subfigure}{0pt}
	\phantomcaption\label{fig:refractive_index_doping:n_real}
	\end{subfigure}
	\begin{subfigure}{0pt}
	\phantomcaption\label{fig:refractive_index_doping:n_imag}
	\end{subfigure}
\includegraphics[scale=0.85]{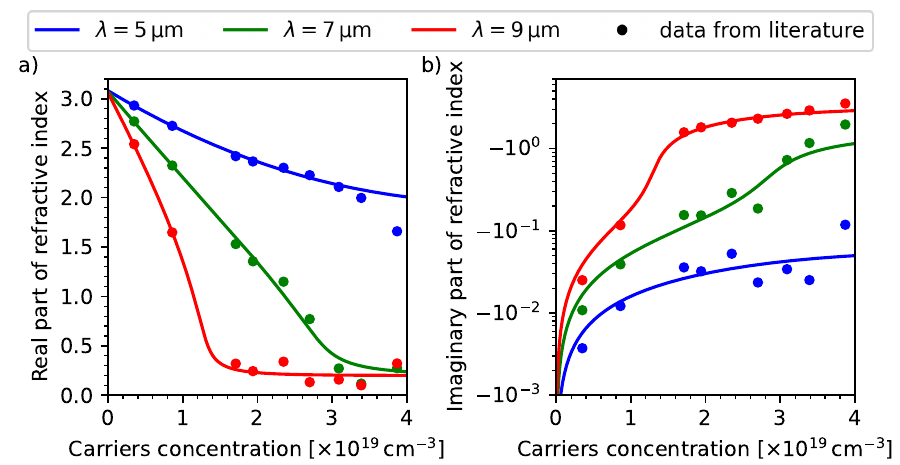}
\caption{a) Real part and b) imaginary part of refractive index of InP. Values from \cite{panah2017} are marked by dots, lines represent dependency obtained by Drude model.}
\label{fig:refractive_index_doping}
\end{figure}

\begin{figure}[htbp]
\centering
	\begin{subfigure}{0pt}
	\phantomcaption\label{fig:refractive_index_wavelength:n_real}
	\end{subfigure}
	\begin{subfigure}{0pt}
	\phantomcaption\label{fig:refractive_index_wavelength:n_imag}
	\end{subfigure}
	\begin{subfigure}{0pt}
	\phantomcaption\label{fig:refractive_index_wavelength:reflectance}
	\end{subfigure}
\includegraphics[scale=0.85]{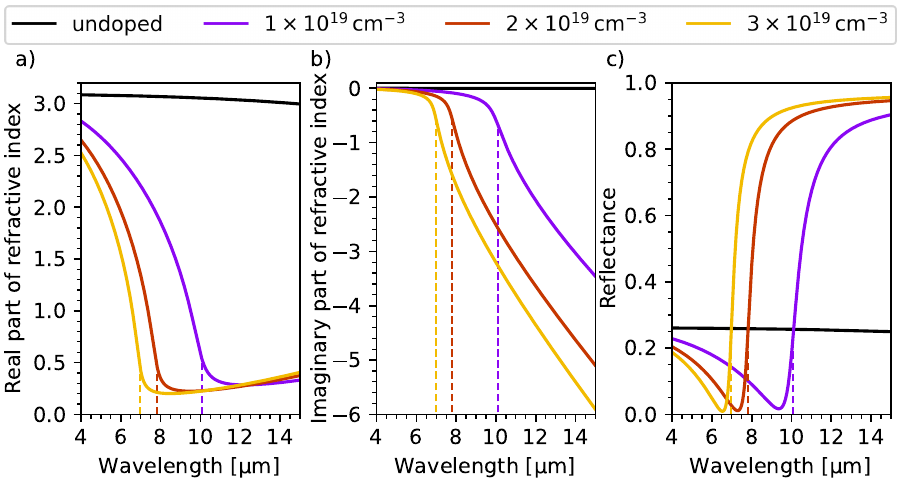}
\caption{a) Real part and b) imaginary part of the refractive index of InP based on experimental characteristics from \cite{panah2017}, and c) reflectivity of a uniform InP layer when illuminated with infrared light incident from air side, all as a function of wavelength for different levels of carrier concentration indicated by colours. Vertical dashed lines indicate plasma frequency.}
\label{fig:refractive_index_wavelength}
\end{figure}

The reflection from the interface between two layers with different complex refractive indices can be described by the Fresnel equations. For normal incidence, the optical power reflectance is given by:
\begin{equation}\label{eq:fresnel}
R = \frac{(n_{2} - n_{1})^2+(k_{2} - k_{1})^2}{(n_{2} + n_{1})^2+(k_{2} + k_{1})^2}
\end{equation}
where $n-\mathrm{i}k$ denotes the complex refractive index, and the respective subscripts refer to materials 1 and 2. The formula \eqref{eq:fresnel} shows that differences in both the real and imaginary parts of the complex refractive index contribute to the reflection. In the mid-infrared range, the frequency of light is close to the eigenfrequency of free electrons in semiconductors, leading to a significant variation in their complex refractive index, which can be described using Drude's model. Figures \ref{fig:refractive_index_doping:n_real} and \ref{fig:refractive_index_doping:n_imag} show the complex refractive index of InP as a function of carrier concentration for different wavelengths. Dots correspond to experimental data reported in \cite{panah2017}, while the lines represent fitted dependencies according to the Drude model. Figures \ref{fig:refractive_index_wavelength:n_real} and \ref{fig:refractive_index_wavelength:n_imag} present the complex refractive index of InP as a function of wavelength for various levels of n-type doping. It can be observed that, in accordance with the Drude model, the real part of the refractive index decreases with increasing wavelength and approaches a minimum close to zero near the plasma resonance wavelength $\lambda_\mathrm{p}$. For wavelengths longer than $\lambda_\mathrm{p}$, the semiconductor enters the plasmonic regime, characterized by a metal-like optical response. In this regime, the dielectric permittivity satisfies the condition $\operatorname{Re}[\varepsilon] = \operatorname{Re}\!\left[(n - \mathrm{i}k)^2\right] < 0 .$ ~\cite{Tommasi2013}. As a result, when infrared light with a wavelength exceeding $\lambda_\mathrm{p}$ is incident on a highly doped InP layer (from the air side in our example see Fig.~\ref{fig:refractive_index_wavelength:reflectance}), plasmonic behavior is observed; however, a high reflectivity is achieved only for wavelengths sufficiently longer than $\lambda_\mathrm{p}$. For sufficiently high doping levels, the reflectance can exceed \pc{90} and saturates at approximately \pc{95} for wavelengths well beyond $\lambda_\mathrm{p}$. The calculations presented in Fig.~\ref{fig:refractive_index_wavelength:reflectance} were performed for an infinitely thick layer of highly doped InP, but similar levels of reflectivity can be achieved for a layer of thickness greater than half the wavelength. However, at wavelengths below $\lambda_\mathrm{p}$, the reflection is less than the level of Fresnel reflection calculated for undoped InP. The calculations of reflection from a homogeneous plasmonic layer presented in this section serve as a reference benchmark for the subsequent analyses in this work.

\section{Numerical optimisation}
\label{sec:optimization}
We utilize the Plane Wave Admittance Method \cite{dems2005} to perform numerical calculations of the optical power reflectance and light distribution in PE DBRs. To optimize the design, we combine an algorithm for calculating reflection with a multidimensional Nelder-Mead simplex algorithm \cite{nelder1965, wright1996}. Our calculations assume a free carrier concentration of \percmcube{2e19} in the doped layers and no free electrons in the undoped layers of the PE DBR . It should be noted that undoped InP typically exhibits a residual free carrier concentration on the order of \num{e13}--\percmcube{e14}; this issue will be further discussed in Section~\ref{sec:experimental_characterisation}. The dispersion of the complex refractive index follows the model depicted in Fig.~\ref{fig:refractive_index_doping} and \ref{fig:refractive_index_wavelength}.
\begin{figure}[htbp]
\centering
	\begin{subfigure}{0pt}
	\phantomcaption\label{fig:fields_optimization:trajectory}
	\end{subfigure}
	\begin{subfigure}{0pt}
	\phantomcaption\label{fig:fields_optimization:quarter_dbr}
	\end{subfigure}
	\begin{subfigure}{0pt}
	\phantomcaption\label{fig:fields_optimization:u_pe_dbr}
	\end{subfigure}
	\begin{subfigure}{0pt}
	\phantomcaption\label{fig:fields_optimization:nu_pe_dbr}
	\end{subfigure}
\includegraphics[scale=0.85]{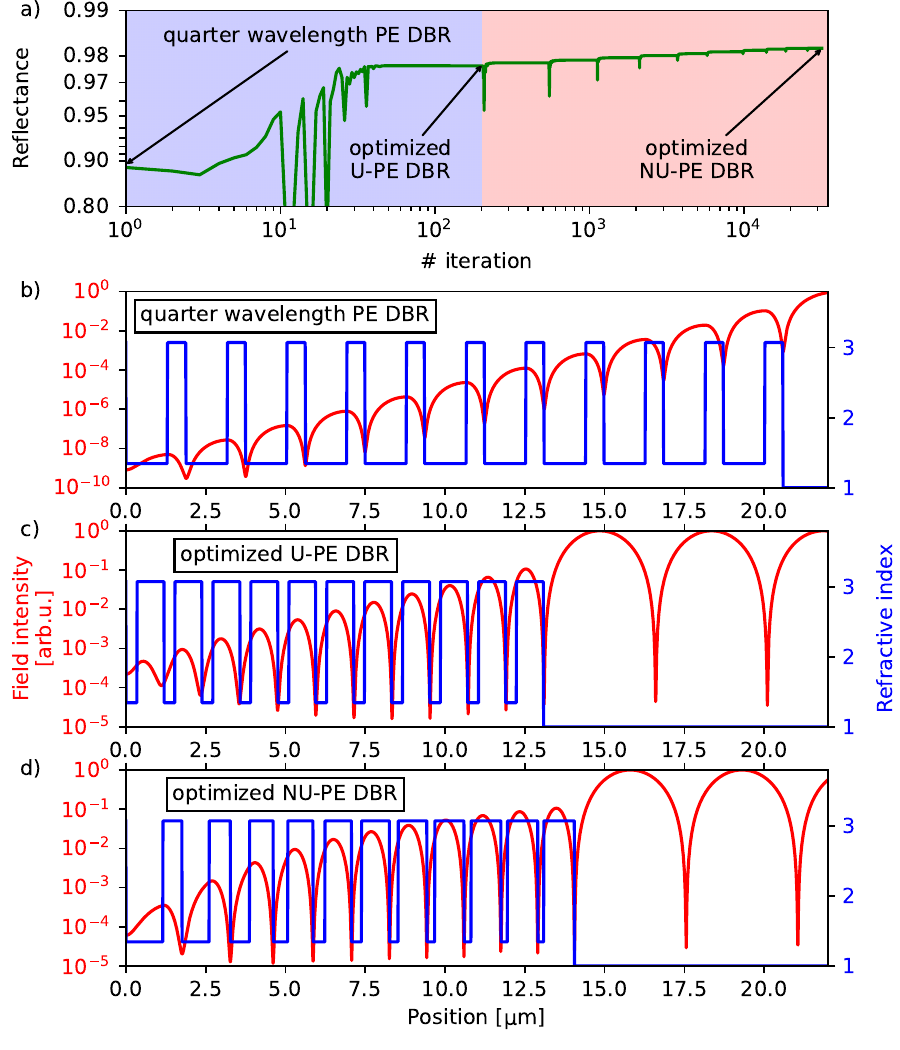}
\caption{a) Optimisation trajectory illustrating the evolution of reflectivity at the wavelength of \um{7} for U-PE DBR and NU-PE DBR as a function of the number of iterations in the optimisation process with variable layer thicknesses. The initial PE DBR configuration assumes quarter-wavelength layer thicknesses. In the first stage (light-blue background), all DBR sections are identical, and the algorithm searches for the optimal thicknesses of the undoped and highly doped layers, resulting in the U-PE DBR design. In the second stage (light-pink background), the optimal U-PE DBR is used as the starting point, and the thickness of each individual layer is varied independently, leading to the optimised NU-PE DBR structure. Panels b), c) and d) show light intensity (red) in logarithmic scale and real refractive index (blue); b) depicts the initial PE DBR configuration, based on the quarter-wavelength layer thickness assumption, serving as the starting point for the optimization process; c) presents the U-PE DBR configuration obtained within first step of optimization procedure, d) presents the NU-PE DBR configuration obtained as the result of the optimization procedure.}
\label{fig:fields_optimization}
\end{figure} 

\begin{figure}[htbp]
\centering
	\begin{subfigure}{0pt}
	\phantomcaption\label{fig:spectrums_25DBR:u_pe_dbr}
	\end{subfigure}
	\begin{subfigure}{0pt}
	\phantomcaption\label{fig:spectrums_25DBR:nu_pe_dbr}
	\end{subfigure}
	\begin{subfigure}{0pt}
	\phantomcaption\label{fig:spectrums_25DBR:reflectance}
	\end{subfigure}
	\begin{subfigure}{0pt}
	\phantomcaption\label{fig:spectrums_25DBR:absorption}
	\end{subfigure}
\includegraphics[scale=0.85]{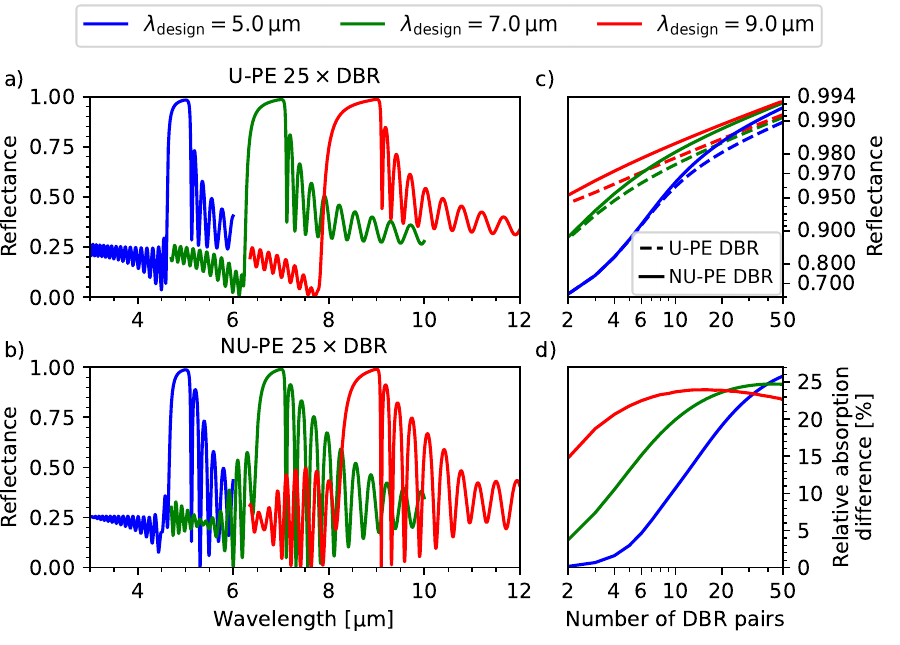}
\caption{Calculated reflection spectra of optimised PE DBRs at the central wavelengths of 5, 7 and \um{9} indicated by different colours in the case of a) U-PE DBR and b) NU-PE DBR; c) maximal reflectance of U-PE and NU-PE DBRs designed for the various wavelengths and d) difference in relative absorption between optimised NU-PE DBR and U-PE DBR versus number of DBR sections.}
\label{fig:spectrums_25DBR}
\end{figure}

In this example, we aim to optimize the configuration of the U-PE DBR composed of 25 DBR pairs to achieve maximal reflectivity at three chosen wavelengths of 5, 7, and \um{9}, using the conventional quarter-wavelength-like design as a starting point. The exemplary optimization trajectories of the U-PE DBR at the wavelength of \um{7}, illustrated in Fig.~\ref{fig:fields_optimization:trajectory}, exhibit minor decaying oscillations as the reflectivity approaches convergence, originating from successive hopping around locally optimal layer thicknesses. The transition from the initial configuration (Fig.~\ref{fig:fields_optimization:quarter_dbr}) to the optimized U-PE DBR (Fig.~\ref{fig:fields_optimization:u_pe_dbr}) is driven primarily by a reduction in the thickness of the highly doped layers and their relocation toward the standing-wave node inside the DBR structure. The three optimised designs of U-PE DBR composed of 25 pairs result in reflectivities of \pc{98.74}, \pc{98.98}, \pc{99.05} at the wavelengths of 5, 7, and \um{9}, respectively, as shown in Figs. \ref{fig:spectrums_25DBR:u_pe_dbr}, \ref{fig:spectrums_25DBR:reflectance}. The layers thickness of optimal configurations are \um{0.600}, \bfum{0.229} for \um{5}, \um{0.941}, \bfum{0.211} for \um{7} and \um{1.281}, \bfum{0.201} for \um{9} (doped layers are in bold). Similar convergence is observed for different numbers of DBR periods. The dependence of the maximum reflectivity on the number of U-PE DBR pairs, resulting from the optimization, is shown in Fig.~\ref{fig:spectrums_25DBR:reflectance} with dashed lines. An increase in the DBR reflectance maximum increases with wavelength which is due to the increasing difference in complex refractive indices between doped and undoped layers. 

In the U-PE DBR, radiation intensity decays exponentially with increasing penetration depth, causing the strongest absorption to occur in the initial layers on the light-incidence side. To reduce absorption, we introduce NU-PE DBR, in which the thickness of each layer in every pair is optimized, using the previously optimized U-PE DBR as the starting configuration. The convergence process, based on the Nelder-Mead simplex algorithm, begins from the optimized U-PE DBR point shown in Fig.~\ref{fig:fields_optimization:u_pe_dbr}. The optimization yields nonlinear dependencies in the thickness of the doped and undoped layers across successive DBR sections, with the doped layers becoming progressively thicker and the undoped layers thinner for DBR sections where light intensity decreases. Interestingly, the thicknesses of the layers within successive sections do not depend on the total number of sections in the NU-PE DBR but only on the section index itself. Consequently, for a given wavelength, there exists a unique and well-defined layer-thickness scheme that is independent of the overall number of sections composing the NU-PE DBR. As shown in Fig.~\ref{fig:spectrums_25DBR:reflectance}, the NU-PE DBRs, indicated by solid lines, exhibit higher reflectivity compared to the U-PE DBRs, demonstrating effective suppression of free-carrier absorption. This effect becomes more pronounced in NU-PE DBRs designed for longer wavelengths. Moreover, their reflectivity increases with the number of DBR pairs, approaching \pc{100}. Figure~\ref{fig:spectrums_25DBR:nu_pe_dbr} illustrates the reflection spectra of NU-PE DBRs for the design wavelengths of 5, 7, and \um{9}, which differ from U-PE DBRs, with a greater slope near the maximum reflectance and a narrower reflection stopband. The comparison of absorption in U-PE and NU-PE DBRs shown in Fig.~\ref{fig:spectrums_25DBR:absorption} demonstrates that the NU-PE structure enables a reduction of nearly \pc{25} in total absorption relative to the U-PE DBR. For PE DBRs with a smaller number of sections, this improvement becomes particularly pronounced at longer wavelengths, indicating a stronger beneficial impact of the NU-PE design compared to the U-PE DBR in this spectral range.

\begin{figure}[htbp]
\centering
	\begin{subfigure}{0pt}
	\phantomcaption\label{fig:other_DBR:gasb_alsb}
	\end{subfigure}
	\begin{subfigure}{0pt}
	\phantomcaption\label{fig:other_DBR:inas_alsb}
	\end{subfigure}
	\begin{subfigure}{0pt}
	\phantomcaption\label{fig:other_DBR:gasb_inas}
	\end{subfigure}
\includegraphics[scale=0.85]{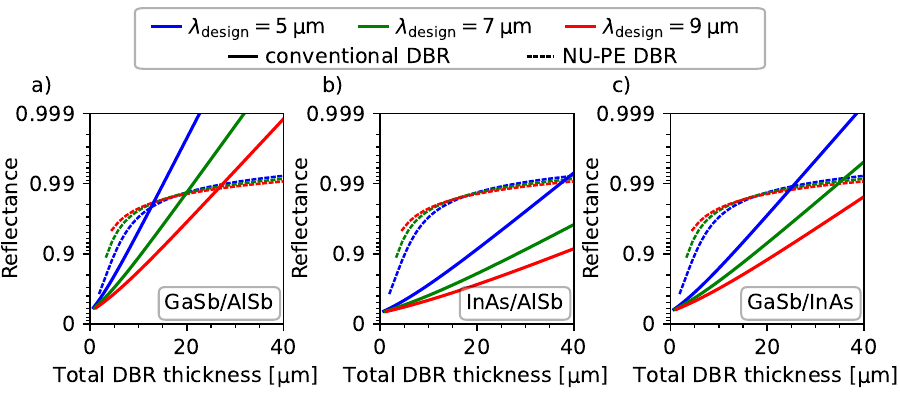}
\caption{Calculated reflectance as a function of total DBR thickness for three design wavelengths (\um{5}, \um{7}, \um{9}) for conventional quarter-wavelength: a) GaSb/AlSb DBR, b) InAs/AlSb DBR, and c) GaSb/InAs DBR. The dashed line in each panel represents the reflectance of the NU-PE DBR for comparison.}
\label{fig:other_DBR}
\end{figure}

Figure~\ref{fig:other_DBR} presents a comparison of the reflectivity of NU-PE DBRs with various configurations of conventional DBRs that could potentially be used in mid-infrared applications, plotted as a function of their total thickness. For thicknesses exceeding \um{10}, the NU-PE DBR exhibits reflectivity that becomes nearly independent of wavelength, resulting in very similar reflectance values for a given total thickness. This behavior makes the NU-PE DBR particularly well suited for longer-wavelength applications. In contrast, conventional DBRs show a much steeper increase toward \pc{100} reflectivity; however, in their case, lattice mismatch leads to cumulative strain in thick quarter-wave layers, which causes relaxation and defect formation, rendering such geometries impractical. As a result, high-quality reflectors in these material systems at shorter wavelengths are typically implemented as ternary-composition materials \cite{sanchez2013}, while in principle they could also be realized as short-period, strain-balanced superlattices \cite{marchewka2023}, rather than as thick individual layers \cite{gossmann1989}. This approach, however, would reduce the effective refractive-index contrast, necessitating thicker stacks to achieve the desired reflectivity.

In both designs of PE DBRs, the light that is not reflected is almost completely absorbed, unlike in conventional DBRs, where the unreflected light is transmitted and only small portion of light is absorbed. Thus, the application of PE DBRs may be restricted to the role of reflecting mirrors without the possibility of simultaneous efficient transmission. Furthermore, PE DBRs exhibit a distinct difference in their reflectance spectrum compared to conventional DBRs, as there is no clear plateau. This is attributed to the significant absorption of radiation at wavelengths different from the design wavelength of the structure.

\section{Experimental characterisation}
\label{sec:experimental_characterisation}
To experimentally validate the optimization results, we fabricated three NU-PE DBRs designed at wavelengths of 5, 7, and \um{9}, respectively. To maintain similar growth conditions for each NU-PE DBR, we grew the DBRs with a total thickness of approximately \um{14}, resulting in 16, 11, and 8 DBR sections designed at wavelengths of 5, 7, and \um{9}, respectively. The simulated values of peak reflectivity and the corresponding full widths at half maximum (FWHM) of the reflection spectra for the three consecutive wavelengths are presented in Table~\ref{tab:spectrums_comparation_data}. The epitaxial growth of alternating n\textsuperscript{++}InP:Si and undoped InP layers was carried out using a low-pressure metalorganic vapor phase epitaxy (LP MOCVD) system by AIXTRON with CCS 3$\times$2" FT reactor. The growth rate remained stable within the range of \num{1.16}--\SI{1.19}{\nano\meter/\second}, independent of both the doping concentration and the layer position. The doping concentration of n\textsuperscript{++}InP:Si was set to \percmcube{2e19}, which represents the maximum level at which crystallographic defects do not compromise the interface uniformity and, consequently, the DBR reflectivity. Further increase of the doping level may be achieved by optimizing the growth temperature to suppress defect formation and mitigate the associated degradation of PE DBR reflectivity. 

\begin{figure}[htbp]
\centering
	\begin{subfigure}{0pt}
	\phantomcaption\label{fig:refractometry_9:all}
	\end{subfigure}
	\begin{subfigure}{0pt}
	\phantomcaption\label{fig:refractometry_9:2nd}
	\end{subfigure}
	\begin{subfigure}{0pt}
	\phantomcaption\label{fig:refractometry_9:8th}
	\end{subfigure}
\includegraphics[scale=0.85]{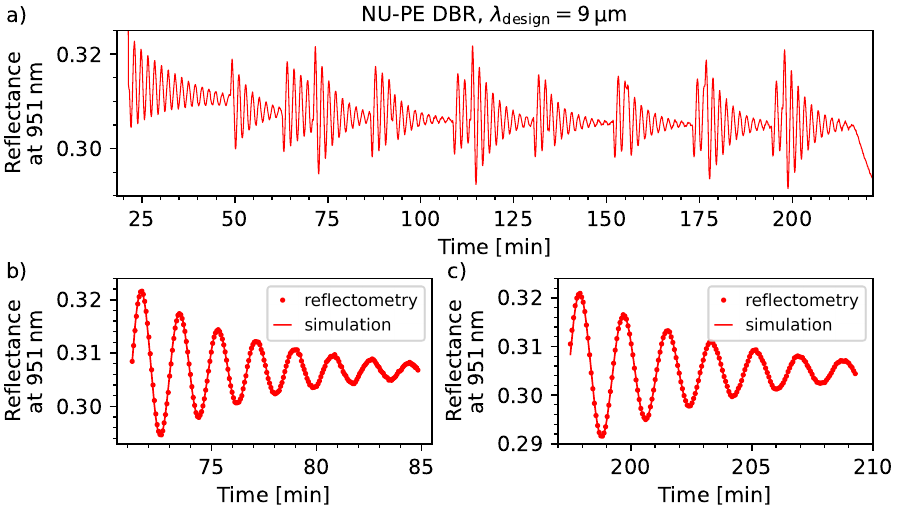}
\caption{In-situ reflectivity at 951 nm recorded during epitaxial growth a) of the complete NU-PE DBR designed for a target wavelength of \um{9}, and comparison of the reflectivity corresponding to the b) 2nd and c) 8th undoped InP layer in NU-PE DBR.}
\label{fig:refractometry_9}
\end{figure}

The growth parameters were monitored using the EpiTT LayTec in situ reflectometry system \cite{laytec}. The reflectometry results for the NU-PE DBR designed for \um{9} are shown in Fig.~\ref{fig:refractometry_9:all}. Variations in reflectivity during the growth process enables precise monitoring of the growth rates and, consequently, the determination of individual layer thicknesses. Figure~\ref{fig:refractometry_9:2nd} presents both simulated and experimental reflectivity for the growth of the second undoped InP layer, as well as the final one (Figure~\ref{fig:refractometry_9:8th}) of the NU-PE DBR. The stable average signal level and constant oscillation period clearly indicate that the surface quality remains unchanged throughout the NU-PE DBR growth, despite the substantial total thickness of \um{14}. These results confirm that, within the homoepitaxial growth regime, the total thickness of the deposited layers does not impose a limitation that would compromise the surface morphology quality of semiconductor structures.
\begin{figure}[htbp]
\centering
	\begin{subfigure}{0pt}
	\phantomcaption\label{fig:comparation_SIMS_SCM:5um}
	\end{subfigure}
	\begin{subfigure}{0pt}
	\phantomcaption\label{fig:comparation_SIMS_SCM:7um}
	\end{subfigure}
	\begin{subfigure}{0pt}
	\phantomcaption\label{fig:comparation_SIMS_SCM:9um}
	\end{subfigure}
\includegraphics[scale=0.85]{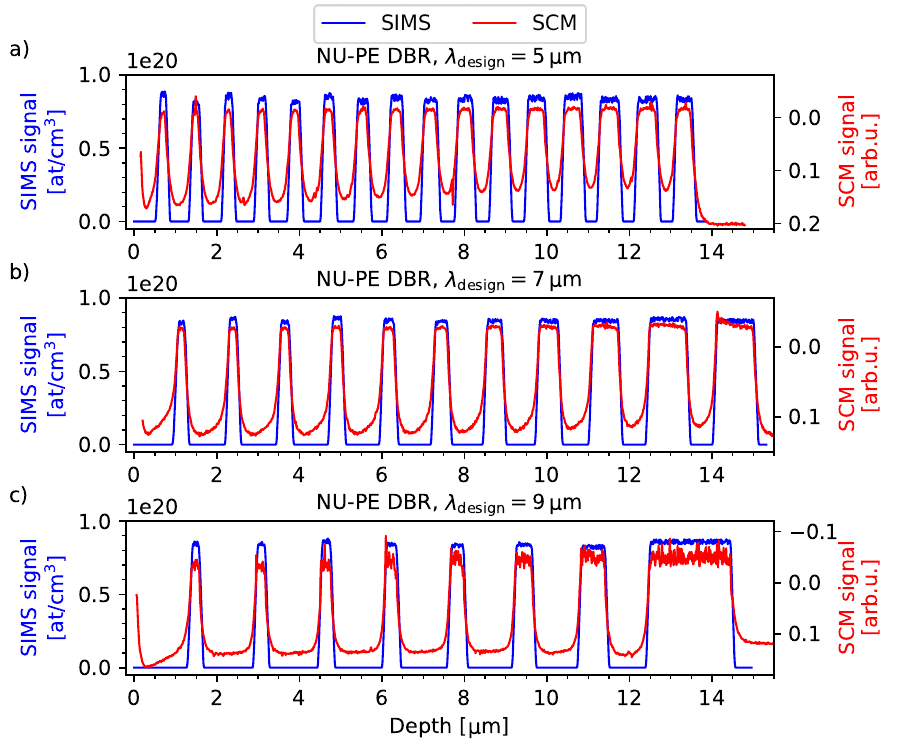}
\caption{Secondary ion mass spectrometry (SIMS) and scanning capacitance microscopy (SCM) profiles of NU-PE DBRs designed at the wavelengts: a) \um{5}, b) \um{7} and c) \um{9}.}
\label{fig:comparation_SIMS_SCM}
\end{figure}

Figure~\ref{fig:comparation_SIMS_SCM} presents profiles measured using secondary ion mass spectrometry (SIMS, blue curves) and scanning capacitance microscopy (SCM, red curves) across the NU-PE DBRs designed for central wavelengths of \um{5}, \um{7}, and \um{9}. In both methods, determining the position of a given layer requires knowledge of the etching rate used in SIMS and SCM. Therefore to accurately estimate the total thickness of the DBRs, we incorporated reflectometry data, which provides the layer thicknesses in the fabricated NU-PE DBRs. SIMS reveals the intended dopant ions distribution with sharp interfaces, whereas SCM reflects the electrically active dopant profile. The smoother transitions observed in SCM between layers of different doping levels may stem from partial dopant activation or compensation effects. These discrepancies indicate that the actual electrical response deviates from the idealized sharp modulation and may therefore influence the effective complex refractive index distribution. 

\begin{figure}[htbp]
\centering
	\begin{subfigure}{0pt}
	\phantomcaption\label{fig:spectrums_comparation:5um}
	\end{subfigure}
	\begin{subfigure}{0pt}
	\phantomcaption\label{fig:spectrums_comparation:7um}
	\end{subfigure}
	\begin{subfigure}{0pt}
	\phantomcaption\label{fig:spectrums_comparation:9um}
	\end{subfigure}
\includegraphics[scale=0.85]{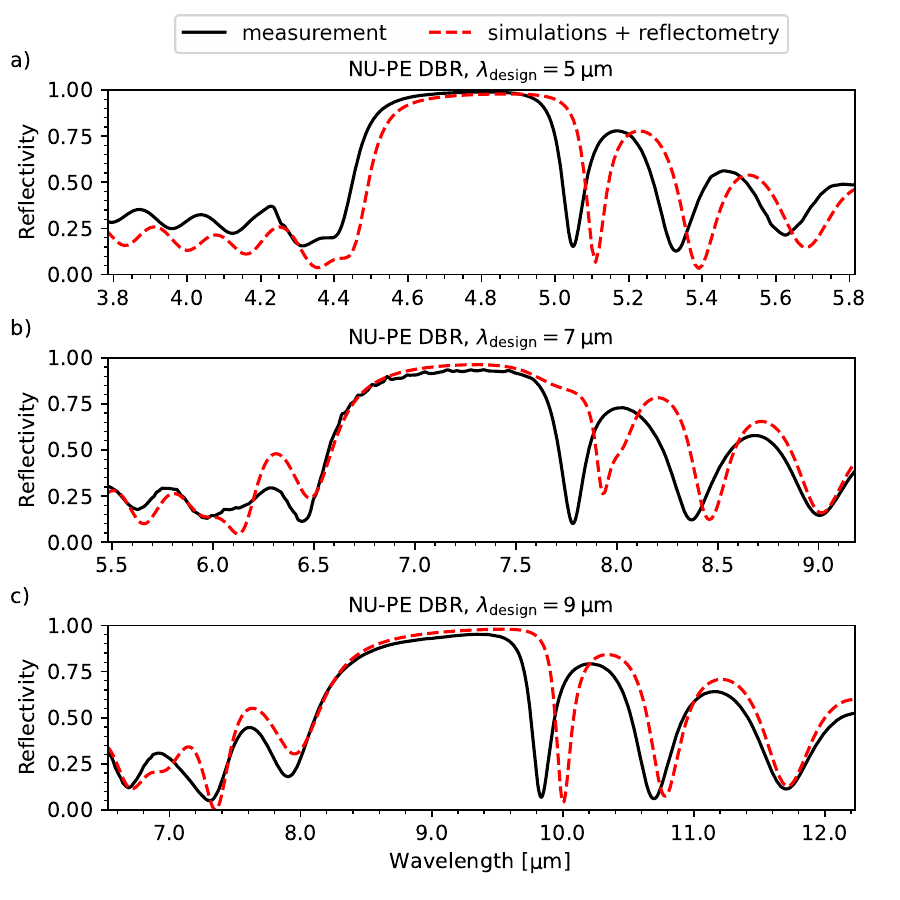}
\caption{Measured (solid lines) and calculated (dashed lines) reflection spectra of the experimentally realized NU-PE DBRs for thicknesses of a) \um{5}, b) \um{7}, and c) \um{9}. The layer thicknesses used in the simulations were taken from the reflectometry measurements.}
\label{fig:spectrums_comparation}
\end{figure}

The reflectance spectrum of the NU-PE DBRs was measured with a Fourier transform infrared spectroscopy (FTIR) setup calibrated against a commercial gold mirror with a spot size of \um{250}. The incident light beam was directed at an angle of 13 degrees with respect to the normal direction of the sample. The reflection from the gold layer was normalized using the Fresnel equation, assuming the complex refractive index of gold from \cite{ordal1987}. 
In Fig.~\ref{fig:spectrums_comparation} the solid lines show the experimental reflection spectra of the NU-PE DBR samples with a stopbands centered at approximately 5, 7, and \um{9}. Maximal reflections and corresponding full-width at half maximum of the reflection spectra of fabricated samples are collected in Table~\ref{tab:spectrums_comparation_data}.

\begin{table}[htbp]
\centering
\caption{Maximum reflectivity and full width at half maximum (FWHM) of the reflection spectra obtained from numerical modelling for the designed NU-PE DBRs at incidence angles of \deg{0} and \deg{13}, along with experimental values measured for the fabricated NU-PE DBRs.}
\label{tab:spectrums_comparation_data}
\begin{tabular}{*{8}{M{1.5cm}}}
\toprule								
&	&	\multicolumn{2}{c}{theory (\deg{0})}	&	\multicolumn{2}{c}{theory (\deg{13})}	&	\multicolumn{2}{c}{experiment}		\\ \midrule
Design wavelength&	Number of DBR pairs&	Maximal reflectivity&	FWHM [\si{\micro\meter}]&	Maximal reflectivity&	FWHM [\si{\micro\meter}]&	Maximal reflectivity&	FWHM [\si{\micro\meter}]	\\ \midrule
\um{5}&	16&	\pc{98.06}&	0.62&	\pc{98.00}&	0.61&	\pc{98.98}&	0.57	\\
\um{7}&	11&	\pc{98.22}&	1.15&	\pc{97.61}&	1.11&	\pc{93.47}&	1.15	\\
\um{9}&	8&	\pc{98.18}&	1.75&	\pc{97.89}&	1.73&	\pc{95.20}&	1.66	\\ \bottomrule
\end{tabular}
\end{table}

The dashed lines in the Fig.~\ref{fig:spectrums_comparation} present the calculated reflectance spectra for light incident at \deg{13}, matching the FTIR measurement conditions. The layer thicknesses used in the simulations were taken from reflectometry results. The refractive index of heavily doped InP was calculated for a carrier concentration of \percmcube{2e19} , using the model shown in Fig.~\ref{fig:refractive_index_doping}. Notably, the spectra shown in Fig.~\ref{fig:spectrums_comparation} were obtained without any further modification of the parameters in numerical model.
A peak reflectance exceeding \pc{90} is reproducibly achieved in the experiment, with a stopband FWHM of at least \pc{11} of the designed wavelength. However, in the case of $\lambda=\um{7}$ the maximum reflectance is approximately \pc{5} lower than the theoretically predicted value. This deviation can be attributed to several factors, including an unintentional background doping level estimated at around \percmcube{7.7e13} imperfect control of high doping concentrations (confirmed by the shape of SCM profile, see Fig.~\ref{fig:comparation_SIMS_SCM}), diffusion of free carriers into nominally undoped regions, possible deviations in layer thicknesses, and calibration uncertainty related to the absolute reflectance of the reference gold mirror.

\begin{figure}[htbp]
\centering
	\begin{subfigure}{0pt}
	\phantomcaption\label{fig:led_spectrums:camera1}
	\end{subfigure}
	\begin{subfigure}{0pt}
	\phantomcaption\label{fig:led_spectrums:camera2}
	\end{subfigure}
	\begin{subfigure}{0pt}
	\phantomcaption\label{fig:led_spectrums:5um}
	\end{subfigure}
	\begin{subfigure}{0pt}
	\phantomcaption\label{fig:led_spectrums:7um}
	\end{subfigure}
	\begin{subfigure}{0pt}
	\phantomcaption\label{fig:led_spectrums:9um}
	\end{subfigure}
\includegraphics[scale=0.85]{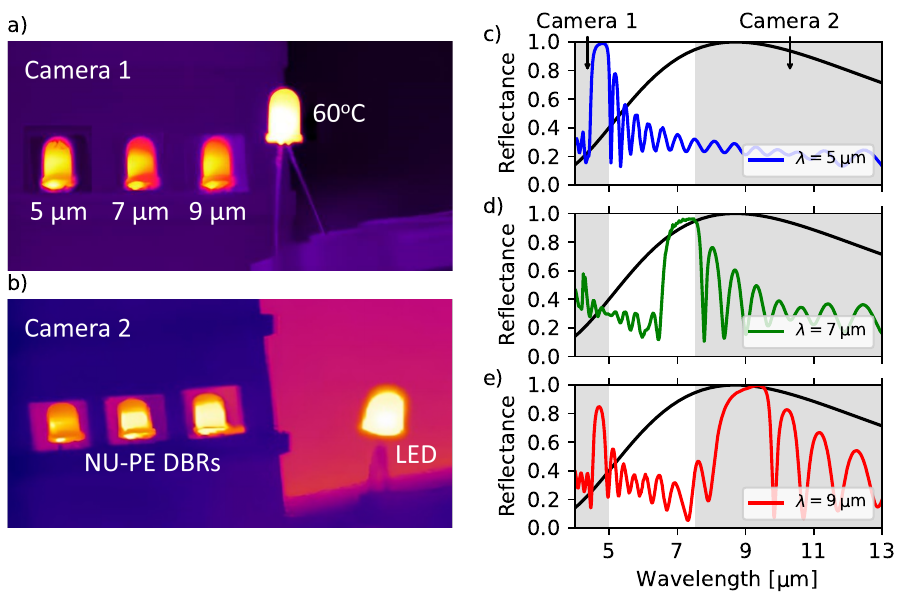}
\caption{Infrared imaging of NU-PE DBRs illuminated by a \degC{60} black-body source. a) Image recorded with the InSb-cooled infrared camera (1.5--\um{5}). b) Image captured with the microbolometer-based camera (7.5--\um{14}). Measured reflectance spectra of the NU-PE DBRs designed for a central wavelength of a) \um{5}, b) \um{7}, and c) \um{9} (blue, red, and black curves, respectively), overlaid with the normalized \degC{60} black-body emission spectrum (black line) and the spectral sensitivity ranges of both cameras (shaded grey).}
\label{fig:led_spectrums}
\end{figure}

Figure~\ref{fig:led_spectrums} visualizes the reflection properties of the NU-PE DBRs. Infrared imaging was performed using an InSb cooled infrared camera operating in the \num{3}--\um{5} spectral range (Camera 1, Fig.~\ref{fig:led_spectrums:camera1}) and a vanadium-oxide microbolometer camera sensitive in the \num{7.5}--\um{14} range (Camera 2, Fig.~\ref{fig:led_spectrums:camera2}). Both cameras image the reflection of infrared radiation from the three NU-PE DBRs discussed in this work. Radiation is emitted by an operating light-emitting diode with a surface temperature of approximately \degC{60}. Figure~\ref{fig:led_spectrums:5um} shows the reflection spectra of the NU-PE DBRs overlaid with the normalized black-body emission spectrum at \degC{60} and the spectral sensitivity ranges of both cameras. The strong reflected signal in Fig.~\ref{fig:led_spectrums:camera1} observed for the NU-PE DBRs designed at \um{5} (blue curve in Fig.~\ref{fig:led_spectrums:5um}) and the \um{9} (red curve in Fig.~\ref{fig:led_spectrums:9um}) arises from the overlap of their reflection maxima with the sensitivity band of Camera 1 (secondary reflection maximum in the case of \um{9} NU-PE DBR). In the case of the \um{7} NU-PE DBR (green curve), the thermogram is dominated by a weak radiation refection in the Camera 1 sensitivity band. In Fig.~\ref{fig:led_spectrums:camera2}, the detected radiation intensity decreases systematically from the long-wavelength to the short-wavelength NU-PE DBRs, as the \um{9} NU-PE DBR coincides with the maximum of the black-body emission and the sensitivity range of Camera 2, while the other two NU-PE DBRs exhibit progressively weaker spectral overlap with the detector’s response range.

To perform electrical characterization of the PE DBRs, transmission line measurements (TLM) were carried out on mesa structures of \um{9} NU-PE DBR defined using standard photolithography and wet chemical etching. The specific vertical resistivity of the PE DBR stack was determined to be $\sim\omcm{1.7}$. This value falls between the reported low resistivity of \omcm{4e-4} characteristic of highly doped n-type InP with an electron concertation of approximately \percmcube{3e19} \cite{pasquariello2006} and the value of \omcm{11.5} measured for intentionally undoped InP \cite{badura2020}. It is also comparable to, and slightly lower than, the calculated resistivities of the PE DBR stacks (\omcm{6.9}, \omcm{7.6}, \omcm{8.0} for \um{5}, \um{7}, \um{9} NU-PE DBRs, respectively) obtained under the assumptions of the above-mentioned layer conductivities and the absence of resistivity-increasing effects such as junction-related barriers or carrier scattering. For reference, conventional n-type arsenide-based DBRs employing modulation doping show resistivities on the order of \omcm{e-3} \cite{wen2020}. Phosphide-based DBRs lattice-matched to InP and designed for VCSELs emitting in the 1.3--\um{1.55} range exhibit resistivities of approximately \omcm{0.4} for InP/InGaAlAs and \omcm{0.8} for AlINAs/InGaAlAs DBRs \cite{lu2004}, the AlP/GaP DBR implemented in a GaP-based LED demonstrates a resistivity of about \omcm{20} \cite{hestroffer2018}. The resistivity of the PE DBRs could be further reduced by introducing intentional doping (up to \percmcube{2e17}) into the originally undoped layers. This would substantially improve the electrical conductivity of the PE DBR, while only minimally affecting the refractive index contrast. Consequently, the reflectivity would slightly decrease, as is typically observed in conventional DBRs. Moreover, in a homoepitaxial PE DBR the absence of interfaces between materials with different bandgaps prevents the formation of band discontinuities and minimizes electron scattering at heterogeneous boundaries, which are inherently present in non-homoepitaxial DBRs.

\section{Conclusions}
\label{sec:conclusions}
We have introduced a novel plasmon-enabled distributed Bragg reflector (PE DBR) architecture for achieving high infrared reflectivity, based on modulation doping in monolithic indium phosphide. Numerical simulations combined with inverse-design optimization demonstrate that a uniform (U-PE DBR) structure composed of 50 periodic pairs of undoped and highly doped (\percmcube{2e19}) layers can achieve reflectivity exceeding \pc{99} for wavelengths above \um{4}. The remaining $\sim\pc{1}$ loss originates primarily from free-carrier absorption, which constitutes the dominant limiting factor. Furthermore, we show that this absorption can be reduced by more than \pc{20} through the implementation of a nonuniform (NU-PE DBR) design, in which the thicknesses of all layers are independently optimized. The optimized NU-PE DBR achieves a reflectivity approaching \pc{99.2}--\pc{99.4} for a 50-pair mirror.

To experimentally verify the performance of the inversely designed NU-PE DBRs, we fabricated homoepitaxial InP mirrors with modulated between unintentional \percmcube{7.7e13} and intentional \percmcube{2e19} Si doping using low-pressure metalorganic vapor phase epitaxy (MOVPE). Each NU-PE DBR, designed for central wavelengths of 5, 7, and \um{9} was grown, with a total thickness of approximately \um{14}. The optical properties were characterized using Fourier-transform infrared (FTIR) spectroscopy, yielding peak reflectance values in the range of 93--\pc{99} and full-width-at-half-maximum (FWHM) bandwidths corresponding to 11--\pc{18} of the design wavelength. A detailed comparison between the simulated and measured reflectance spectra showed qualitative and quantitative agreement, confirming the predictive accuracy of the numerical model. The electrical characterization confirmed the expected behavior of the junction-free homoepitaxial configuration, moreover, the conductivity can, in principle, be modified through low-level carrier doping of the initially undoped layers. Importantly, such an approach would allow for a reduction in resistivity without significantly affecting the refractive index contrast, thereby preserving the high reflectivity of the PE DBRs.
Unlike conventional DBRs, whose performance is constrained by growth time and the critical thickness imposed by strain, the plasmon-enabled DBR could provide high reflectivity over a broader spectral range. The key advantage of plasmon-enabled DBRs lies in the ability to tailor the complex refractive index of the semiconductors through the concentration of free carriers, providing a versatile platform that can be applied to all semiconductor materials. This approach opens up new possibilities for the design and fabrication of high-performance optical devices for a variety of applications, such as lasers, optical filters, and modulators with enhanced performance and functionality.

\section*{Acknowledgements}
The authors thank Fabian Hartmann (Julius-Maximilians-Universität Würzburg) for valuable discussions and support. T.C., M.M. and M.J. acknowledge financial support by the Swiss Contribution to reducing economic and social disparities in the EU and from the state budget through the National Centre for Research and Development, Call 2024/90/2025, project entitled “Quantum-cascade vertical cavity surface emitting laser for gas sensing” (QCVCSEL). This work has been completed while M.J. was the doctoral candidate in the Interdisciplinary Doctoral School at the Lodz University of Technology, Poland.
\bibliographystyle{unsrt} 
\bibliography{bibliography}

\end{document}